# Morphological Image Similarity Search on the ALMA Science Archive Query Interface Using Deep Unsupervised Contrastive Representation Learning

Felix Stoehr[1]
Andrea Farago[1]
Stefan Curiban[1]
Alisdair Manning[1]
Jorge Garcia[2]
Pei-Ying Hsieh[3]
Andrew Lipnicky[4]
Adele Plunkett[4]

[1] ESO
[2] Joint ALMA Observatory, Santiago, Chile
[3] National Astronomical Observatory of Japan, Mitaka, Japan
[4] National Radio Astronomy Observatory, Charlottesville, USA

With the exponential growth of astronomical data over time, finding the needles in the haystack is becoming increasingly difficult. The next frontier for science archives is to enable searches not only on observational metadata, but also on the content of the observations themselves. As a step in this direction, we have implemented morphological image similarity search into the ALMA Science Archive (ASA). To achieve this we use self-supervised contrastive affine-transformation-independent representation learning of source morphologies with a deep neural network. For a given image on the ASA web interface, astronomers are presented with a summary view of the morphologically most similar images. Each time an astronomer selects an additional image from that view, the display is instantly updated to show the images most similar to the combination of the selected images. Each selection thus refines the similarity display according to the scientific needs of the astronomer. This is the first time image similarity search has been offered in an astronomical science archive.

## Introduction

The amount of astronomical data is growing exponentially (for example, Stoehr, 2019) resulting in an exponential increase of data available per astronomer. No longer data but now astronomers are becoming — or have already become — the rare resource in astronomy (Stoehr et al., 2015). For science archives like the Atacama Large Millimeter/submillimeter Array (ALMA) Science Archive (ASA)[1] (Stoehr et al., 2017), this evolution means that the next frontier is to make the actual content of the data searchable (Durand et al., 2017), in addition to the searches offered on observational metadata. ALMA has taken some small initial steps in this direction (Stoehr et al., 2022): users can search for similar abstracts of publications and projects on the ASA query interface using a state-of-the-art text similarity method. Additionally, we offer preview images that show lines identified through the ALMA Data Mining Toolkit (ADMIT; Teuben et al., 2015). In the remainder of this paper we detail our work on image similarity, which, together with the above, fits into the 'fastronomy' concept laid out by Stoehr et al. (2022), emphasising that astronomers need to be able to get results quickly.

## Self-supervised contrastive learning

Methods like transfer learning (for example, Peek et al., 2020) or autoencoders have been used with some success for unsupervised image similarity learning — that is, learning without using human classification (labels). Recently, however, a breakthrough has been achieved such that contrastive unsupervised methods perform as well as, or even better than, supervised methods. Such methods have already been applied successfully to astrophysics. For a summary see Huertas-Company et al. (2023).

Here, we use the SimCLR algorithm (Chen et al., 2020) in the implementation made by da Costa et al. (2022)[2]. SimCLR uses self-supervised learning by applying two transformations to each input image (for example, crop, colour change), running both of them through a deep neural network such as ResNet-50. A contrastive loss function is then applied to both output vectors and back-propagated so that the network will learn to return more similar vectors for those transformations (Figure 1). When the network is given transformations of two different images, it is trained to push the output vectors apart.

Contrastive learning is a very powerful approach, as it can work with any type of data (spectra, images, cubes, etc.), can work with any type of network architecture, is conceptually simple and converges well. The most compelling properties of contrastive learning, however, are that the loss function is independent of the specific problem at hand and that additional physically motivated knowledge — such as rotational symmetry for astronomical images — can be directly and straightforwardly encoded by specifying appropriate transformations.

Several modifications to the original algorithm were necessary. In particular, we have added affine transformations including rotation with arbitrary angles, shear and scale. For the latter, adding scale factors less than one (zooming out) had a very positive effect on the quality of the results. We also removed the striping that is applied to the first layers of ResNet-50 as ResNet-50 typically starts with substantially larger images than those we use. A hyperparameter search was conducted on the minimum and maximum scaling factor (0.4, 2.0), the range for shearing (10 degrees), the image normalisation (1), the batch size (256), the dimensions of the output vector (2048), and the image crop size (48 pixels). The values in parenthesis indicate the chosen settings.

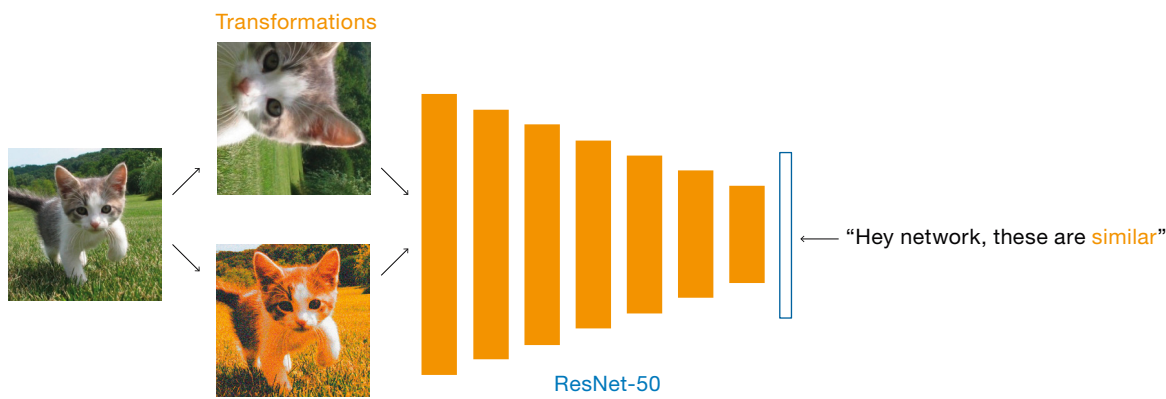

Figure 1. SimCLR principle: A deep neural network architecture is trained by applying two transformations to an input image, running them both through the same network and using a contrastive loss function that will bring the result vectors closer together, while pushing apart vectors of transformations from different input images.

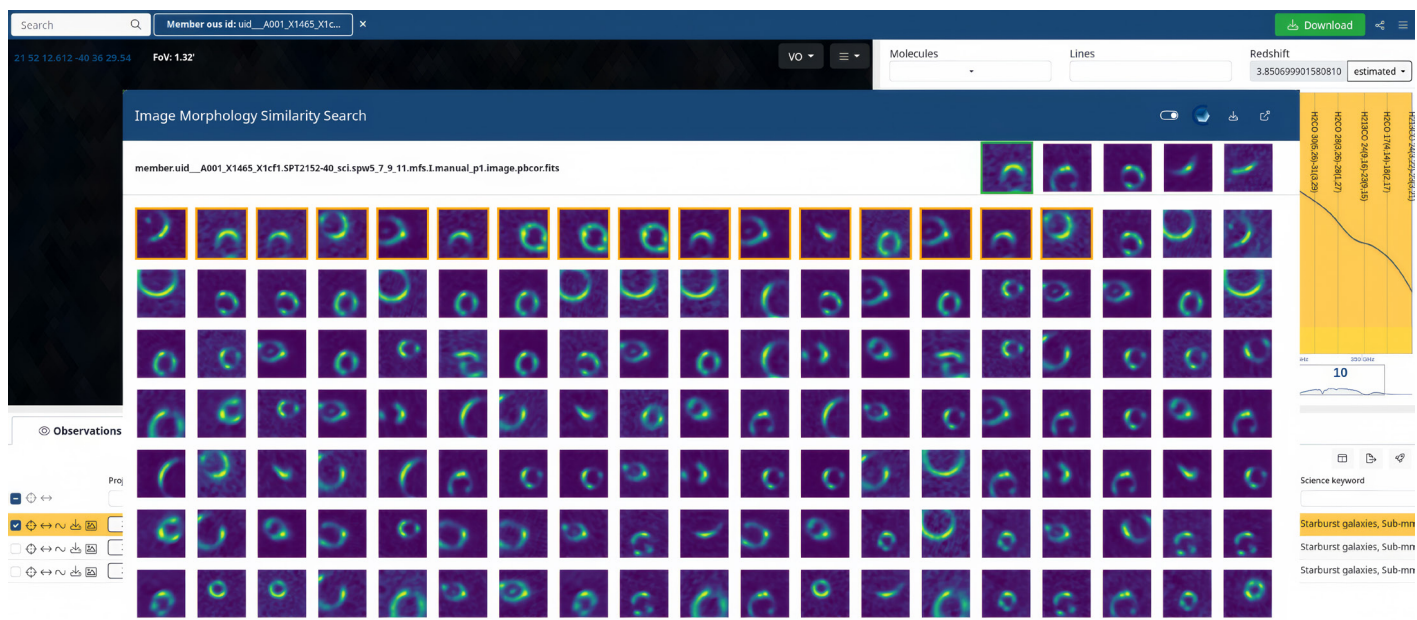

Figure 2. Morphological Similarity Search Interface on the ASA showing the first images that are most similar to the original image in the top left corner. Each additional selection on the interface will trigger a reordering of the remaining images so that they are most similar to the combination of all selected images, thus allowing astronomers to interactively refine the search to their needs.

As images we use cut-outs of 64 × 64 pixels centred on the brightest pixel in the image if the signal to noise is greater than five, and on the centre of the ALMA image otherwise. The image values are scaled to their median with a standard deviation of one. We trained the similarity on all continuum images in the ASA, as well as separately on all the available peak-flux images (sometimes also called 'moment 8') for line cubes. Training the algorithm on 100 000 images for 100 epochs takes about 1.8 hours on three NVIDIA TITAN RTX GPUs. The final evaluation used 10 000 training epochs.

### Results

After the training the output vectors for each image must be inferred and then the pairwise cosine distance of all vectors must be computed to identify the most similar images. To overcome the $O(N^2)$ scaling of this calculation, we normalise the result vectors and can then use the Euclidean distance measurement, thus enabling the use of a kd-tree and reducing the complexity to N log(N). By also using a kd-tree implementation on the GPU[3], which is for our purposes about 60 times faster than running on the CPU, the inference — finding the 1000 images most similar to each of 100 000 images — takes about 1.7 hours on a single GPU.

During this process we also make sure to limit the images that can be listed as similar to a given image to data from a similar scientific category, to avoid scientifically nonsensical similarities, such as between a galactic disc and a protoplanetary disc. We construct the list of scientific categories that are acceptable for an image of a given category by analysing the statistics of scientific categories that have been used together in ALMA publications.

### Caveats

A few astronomers have been asked to validate the usefulness of the similarity for images of their own data. This proved very successful. For example, when given the image of protostellar system HH212, which looks to the eye approximately like an ellipse, the algorithm returns images of similar protostars with outflows, such as BHR71_IRS1 or HOPS-108. Similarly, when given an image of the strongly gravitationally lensed galaxy SPT2145-40, other such lensed galaxies like PJ022643.0 or SPT0402-45 are returned.

However, there are a few caveats to keep in mind. First, there is no unique answer to image similarity. The choice of the hyperparameters does have a substantial influence on the result. Second, there is no objective 'similarity' for astronomical images in the first place. What astronomers need depends strongly on their science case, even for the same initial image. Third, the method can find similar images (analogous to the "You might also like these books" algorithm from online book shops), but it will not necessarily find all similar images. The similarity selections cannot therefore be used for statistical analysis.

Finally, we note that the service is really only based on images. With the heterogeneous nature of ALMA observations in very narrow spectral windows, and the morphologies of the emission from different line transitions varying strongly despite being from the same source and at the same spatial resolution, a *source*-similarity search cannot be built, although this would be desirable.

### Interface

On the ASA query interface[1], astronomers can now find a new icon in the preview pop-ups leading to the image similarity interface (Figure 2), where the most similar images are shown and more and more images are loaded in through virtual scrolling. If the user selects an additional image of interest (indicated by orange frames), the interface combines the vector of the 1000 images most similar to that image with those of all previously selected images on the fly and redisplays the new most similar images in the updated order. By selecting images interactively, astronomers thus can constantly refine the similarity they are seeking, which is crucial given that no objective, one-size-fits-all similarity definition is possible (see also Lochner & Bassett, 2020).

In addition, we have implemented a k-means clustering algorithm to cluster the vectors of the 1000 most similar images in five groups. The most representative image of each cluster is displayed in the top right corner. When the astronomer clicks on an image of this 'quick select' row, the interface pre-selects the 15 images closest to that representative image.

Depending on whether the astronomer starts on the interface with a continuum (total number 259 126 at the time of writing) or a peak-flux image of a line-cube (total number 196 322), we show the most similar continuum or peak-flux images, respectively.

To honour the proprietary period, the similarity interface only displays images of public data. As the inference is very fast to compute, we can construct a new similarity matrix for the query interface regularly to include images that have become public recently. The web interface itself only relies on the pre-computed matrix and makes use of standard web technology. It does not need to be connected to a GPU.

### Outlook

So far the hyperparameters are optimised using only visual inspection. However, in many cases we do have several images for each source in a dataset. This allows us to construct an overall score value from the similarity distance of images that we know are from the same source. We then can run a hyperparameter search to fine-tune parameter setting, minimising the score value automatically. We are also considering collecting the selections users are making on the ASA interface — entirely anonymously — and using those as further input to improve the hyperparameters in future re-trainings.

### Acknowledgements

We gratefully acknowledge discussions with Kai Polsterer, Josh Peek, Martino Romaniello, Vojtech Cvrcek, Abhijeet Borkar, Ashley Barnes, Hannah Stacey, Fabrizia Gluglielmetti, Per Bjerkeli and Leon Boschman. We acknowledge small editorial suggestions provided by a large language model on the final text. ALMA is a partnership of ESO (representing its member states), NSF (USA) and NINS (Japan), together with NRC (Canada), NSTC and ASIAA (Taiwan), and KASI (Republic of Korea), in cooperation with the Republic of Chile. The Joint ALMA Observatory is operated by ESO, AUI/NRAO and NAOJ.

### Links

[1] ASA web interface: https://almascience.org/aq
[2] Solo-learn: https://github.com/vturrisi/solo-learn
[3] cuml, available from https://github.com/rapidsai

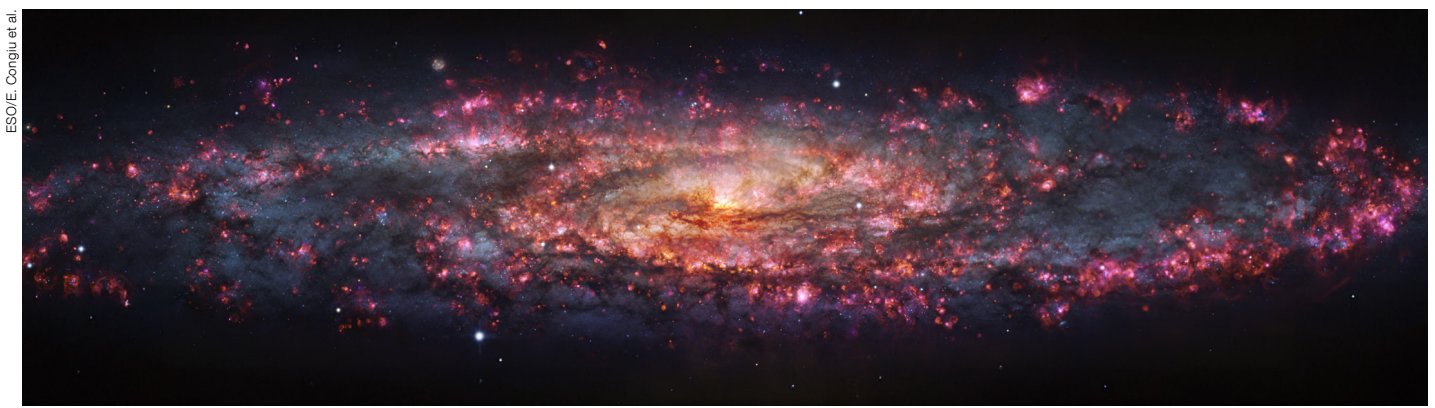

ESO/E. Congiu et al.

This image shows a detailed, thousand-colour image of the Sculptor Galaxy captured with the MUSE instrument at ESO's Very Large Telescope (VLT). Regions of pink light are spread throughout this whole galactic snapshot, which come from ionised hydrogen in star-forming regions. These areas have been overlaid on a map of already formed stars in Sculptor to create the mix of pinks and blues seen here.